\documentclass[twocolumn]{aastex631}

\received{May 3, 2023}
\revised{November 7, 2023}
\accepted{November 19, 2023}
\submitjournal{JAAVSO}

\shorttitle{ASAS-SN RV Tau Variables}
\shortauthors{Nere et al.}
\graphicspath{{./}{figures/}}

\begin{document}

\title{An Audit of the Light Curves of RV Tau Variable Stars in the ASAS-SN Database}
\correspondingauthor{Rachel N. Nere \& Rodolfo Montez Jr.}
\email{rachel.nere@gmail.com,rodolfo.montez.jr@gmail.com}

\author[0009-0000-4358-5561]{Rachel N. Nere}
\affiliation{University of Massachusetts - Boston}

\author[0000-0002-6752-2909]{Rodolfo Montez Jr.}
\affiliation{Center for Astrophysics $\vert$\ Harvard\ \&\ Smithsonian, Cambridge, MA, USA}

\author[0000-0002-0947-569X]{Sophia S\'{a}nchez-Maes}
\affiliation{Center for Astrophysics $\vert$\ Harvard\ \&\ Smithsonian, Cambridge, MA, USA}

\begin{abstract}
RV Tauri variable stars are pulsating evolved stars that are identified by a characterizing feature in their light curves: alternating deep and shallow minima. 
Many RV Tauri variable stars were originally classified decades ago using visual observations and photographic plates. 
However, recent studies suggest that there are imposters, or misclassified variables, amongst the sample of RV Tau stars. 
In this study, we examine 84 known and suspected RV Tau stars that appear in the variable star catalog of the All Sky Automated Survey for Supernova (ASAS-SN). 
For each ASAS-SN light curve, we performed a period analysis and compared our results with those of the General Catalog of Variables Stars (GCVS) and the automatic classification algorithm used by ASAS-SN. 
We found that the pattern of alternating minima present in RV Tau variable stars often confuses automatic classification. 
Our results include updated periods and classifications for our sample.
Our study provides an important step towards obtaining a robust sample of RV Tau variable stars to better understand the pulsation mechanism and the evolutionary pathway of these variable stars.
\end{abstract}


\section{Introduction} \label{sec:intro}

RV Tauri (RV Tau) variable stars are luminous pulsating evolved stars, likely post-asymptotic giant branch (post-AGB), whose variability is characterized by alternating deep and shallow minima. 
RV Tau variable stars have spectral types F-G at maximum light and K-M at
minimum and the supergiant luminosity class \citep{2002PASP..114..689W,2017ARep...61...80S}. 
The formal period, which is the time it takes to complete a deep and shallow cycle, is typically 30-150 days \citep{2017ARep...61...80S} although the upper limit is not very well defined \citep{1991IBVS.3557....1Z}, these stars are classified as {\tt RVA}. 
A subset also exhibit long term variations in their mean magnitude, often with periods of several hundred days, and these stars are classified as {\tt RVB}.
In this work we used the publicly available optical light curves from the All-Sky Automated Survey for Supernova \citep[ASAS-SN,][]{2014ApJ...788...48S,2019MNRAS.486.1907J,2020MNRAS.491...13J} to scrutinize a sample of RV Tau variable stars from the General Catalog of Variable Stars \citep[GCVS,][]{2017ARep...61...80S}. 

\section{Methods} \label{sec:methods}

\subsection{Sample and data}

The stars considered in this work were derived from GCVS \citep{2017ARep...61...80S}, specifically, all the stars with the following RV Tau variable star designations: {\tt RV}, {\tt RVA}, and {\tt RVB}, along with the corresponding uncertain designations: {\tt RV:}, {\tt RVA:}, and {\tt RVB:}.   
We cross-correlated the GCVS-derived list of stars with objects in the ASAS-SN variable star catalog and obtained the light curves from the ASAS-SN variable star database\footnote{https://asas-sn.osu.edu/variables}. 
Our final sample consists of 84 stars, of which, 29 are suspected RV Tau variables, i.e., those with uncertain designations (26 {\tt RV:}, 2 {\tt RVA:}, and 1 {\tt RVB:}), and 55 are known RV Tau variable stars (10 {\tt RV}, 37 {\tt RVA}, and 8 {\tt RVB}).

The ASAS-SN variable star database contains V-band light curves with observations performed between 2013 and 2018 \citep{2019MNRAS.486.1907J}. 
These observations were acquired with telescopes in Hawaii and Chile to a depth of V$\lesssim$ 17\ mag and with a saturation limit of 10-11 mag \citep{2017PASP..129j4502K}.
The number of data points ranges from 69 to 579 with an average of 234 observations per object and span from 2 to 6 years (see Tables \ref{tab:rvpgcvs} and \ref{tab:rvgcvs}). 
18 of the 84 objects have mean V-band magnitudes in the saturation regime (V$<$11 mag). 
However, ASAS-SN applies a correction for saturated stars that produces good light curves for some variable stars in the 10-11 mag range \citep{2019MNRAS.486.1907J}.
Of the 18 stars in the saturation regime, 5 have V$<$10 mag (CE Vir, V Vul, TW Cam, SX Cen, and RV Tau), thus firm conclusions for these objects are not advised. 

\begin{figure*}
\plotone{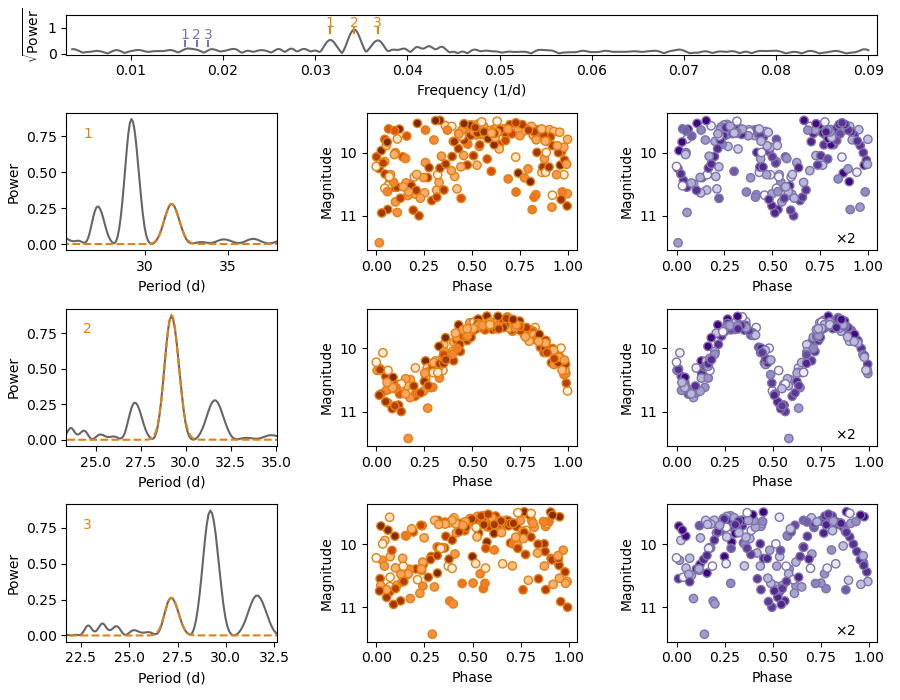}
\caption{An example of a period search with the Lomb-Scargle periodogram for the RVA variable star DI Car. The top panel depicts the square-root of the periodogram power over the entire frequency search range. 
There are three peaks above the false alarm probability threshold that are labeled by the orange numbers. We note that in this case, peaks ``1'' and ``3'' are aliases of peak ``2''. The double-periods are labeled by the purple numbers. The grid of subplots depict the results of fitting the three periodogram peaks. {\it Left} to {\it right}: we depict (i) a zoom-in of the periodogram in period-space and the best-fit Gaussian (orange dashed line), (ii) the phase-folded light curve based on the best-fit period (orange data points), and (iii) the phase-folded light curve based on the double-period (purple data points). 
In the light curves, the color-gradient is correlated with the time of each data point (lighter points are early in the light curve and darker points are later). 
In this example, the double-period peak of periodogram peak ``2'' was selected as the optimal period.   
\label{fig:pgram}}
\end{figure*}

\subsection{Period determination}

The Lomb-Scargle periodograms \citep{1976Ap&SS..39..447L,1982ApJ...263..835S} were calculated for each light curve using the {\tt LombScargle} routine distributed in the {\tt astropy} package \citep{2013A&A...558A..33A,2018AJ....156..123A}. 
We searched for periods between 10 and 275 days and sampled the typical periodogram peak with 10 samples ({\tt samples\_per\_peak=10}). 
We then performed a least-squares fitting of Gaussian curves to the periodogram peaks with a false alarm probability $\geq$5\% using the {\tt curve\_fit} routine from the Scientific Python ({\tt scipy}) package. 
For each peak, we phase-folded the light curve using the best-fit period and double the best-fit period, hereafter referred to as the double-period.  
We selected the optimal period by a visual inspection of the phase-folded light curves.  
Specifically, the optimal period corresponded to the period that produced phase-folded light curves that showed alternating minima and minimal scatter. 

A summary of our analysis for the RVA variable star DI Car can be seen in Figure~\ref{fig:pgram}. 
In this example, three period peaks were found to lie above our false alarm probability threshold and the three best-fit Gaussian curves to the peaks are shown along with the phase-folded light curves for the best-fit periods and their corresponding double-periods. 
In the case of DI Car, the double-period for the highest peak (labeled ``2'' in Figure~\ref{fig:pgram}) was selected as the optimal period because alternating minima are present and there is minimal scatter in the phase-folded light curve; peaks ``1'' and ``3'' are aliases of peak ``2''. 
A peak was identified in every object and in a few cases, more than 10 peaks were identified. 
We examined every peak in order to select the optimal period. 

\begin{figure}
\begin{center}
\plotone{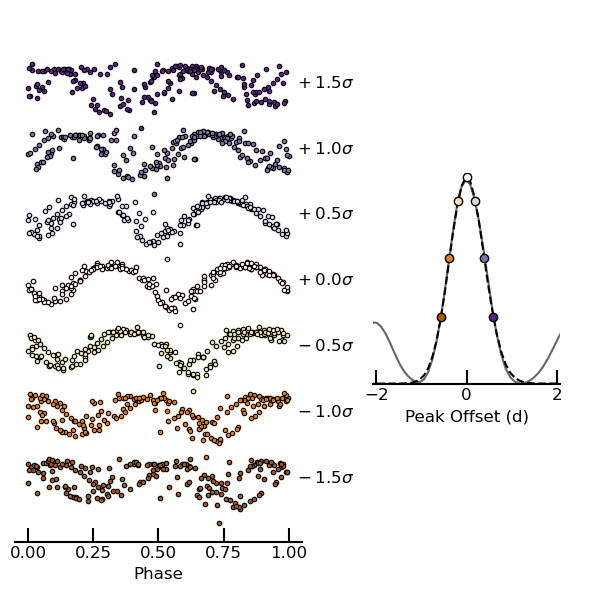}    
\end{center}
\caption{An example of the error analysis of the period search for the RVA variable star DI Car. The left panel depicts phase-folded light curves at offsets of the peak of the periodogram based on the best-fit Gaussian curve. The right panel shows the periodogram peak (solid line), its best-fit Gaussian curve (dashed line), and the offsets used on the left panel. This diagnostic was performed for all periodograms studied in our sample.  
\label{fig:uncertainty}}
\end{figure}

To estimate the uncertainty on the period, we used the best-fit Gaussian curve and its best-fit $\sigma$-value. 
In Figure~\ref{fig:uncertainty} we depict phase-folded light curves from -1.5$\sigma$ to +1.5$\sigma$ to demonstrate that alternating minima can be seen for a range of period values centered on the peak period. 
Based on our review of several stars with short and long periods, we determined that visual inspection of a phase folded light curve would result in the RV Tau classification for periods between -1$\sigma$ and +1$\sigma$, thus we take \textit{the specified RV Taus mentioned in the classification} as the uncertainty on the period and report this in our results. 
There is evidence for period changes amongst some RV Tau stars \citep{1991ApJ...375..691P}, longer term studies are necessary to identify such period changes and the presence of unidentified period changes are likely to result in larger uncertainty in the period determination. 

\subsection{Classification criteria}\label{classcriteria}

\citet{1943AnHar.113....1P} gives three principal characteristics of RV Tau variable stars: (1) alternating minima are deep and shallow, (2) the minima interchange, and (3) the spectrum at maximum is F, G, or K. 
In this study, we only examine the light curves and determine if RV Tau variability is present if any of the following conditions are met:  
\begin{itemize}
    \item alternating minima are present each cycle and the deep and shallow depths are consistent,
    \item alternating minima are present each cycle and the deep and/or shallow depths vary, 
    \item alternating minima are present each cycle and long-period variation of the mean magnitude is present (RVB-phenomenon).
\end{itemize}
In some cases, alternating minima can appear to interchange, we consider these an example of the second criteria listed above, and thus, consistent with RV Tau variability. 
In a few cases, alternating minima were found, but further examination revealed that the light curves were more consistent with binary phenomena (short period binary and/or eclipsing binary systems).
Some light curves contain outliers in their measurements that are explained by ASAS-SN pipeline reduction artifacts \citep[detailed further in][]{2017PASP..129j4502K}. 
These outlier measurements do not significantly impact our period analysis. 
There are some light curves with no apparent RV Tau variability nor apparent periodicity over the range we considered.

\section{Results} \label{sec:results}

We have studied ASAS-SN light curves for 84 objects found in the GCVS and ASAS-SN Variable Star Database catalogs, their phase-folded light curves are presented in Figures~\ref{fig:rvp}-\ref{fig:rva}. 
Our periods and classifications are listed in Tables~\ref{tab:rvpgcvs}-\ref{tab:rvgcvs}. 
For comparison, we also compiled the periods from GCVS and the ASAS-SN Variable Star Database and include them in Tables~\ref{tab:rvpgcvs}-\ref{tab:rvgcvs}. 
The GCVS periods are determined from a variety of methods while the ASAS-SN periods are also determined from Lomb-Scargle periodograms \citep[see][for more details]{2017PASP..129j4502K,2019MNRAS.486.1907J}.
In Tables~\ref{tab:rvpgcvs} and \ref{tab:rvgcvs}, we list the adopted periods in the $P_{\rm Adopted}$ column. Periods with uncertainty values are those derived in this work from our analysis of the periodograms. 
Where our analysis does not
converge on a solution, we provide the best result from our
periodograms, GCVS, or ASAS-SN (including double-periods)
based on visual inspection and omit error bars. Overall, our
periods tend to agree with the periods reported in GCVS and
often differ from the ASAS-SN periods. 

\subsection{Comparison between GCVS and ASAS-SN classifications}

There are 29 objects classified by GCVS as suspected RV Tau, i.e., those with {\tt RV:}, {\tt RVA:}, and {\tt RVB:} designations that have ASAS-SN light curves within the variable star database. 
We identified the alternating minima in 11 of these objects. 
8 remain inconclusive (classified as {\tt U} and {\tt NP} in Tables~\ref{tab:rvpgcvs} and \ref{tab:rvgcvs}) and 10 are unlikely to be RV Tau variables based on the ASAS-SN light curves. 
The ASAS-SN classification algorithm identified none of these as RV Tau variable stars. 
Over 50\% of these stars are identified as semi-regular and irregular variable stars by ASAS-SN (often using {\tt SR} as a generic classification, despite the fact that some are longer period {\tt SRd} variable stars). 
Only two are identified as Cepheid variable stars by ASAS-SN (one as a fundamental mode Type I Cepheid [{\tt DCEP}] and one as a W Virginis Type II Cepheid [{\tt CWA}]). 

There are 55 objects classified in GCVS as {\tt RV}, {\tt RVA}, and {\tt RVB} that have ASAS-SN light curves within the variable star database. 
We identified the clear presence of alternating minima in 43 of these objects. 
There remaining 5 objects exhibit alternating minima but with depths that are irregular. 
In V0967 Cyg, TX Per, and TW Aql, the alternating minima seem to disappear for some cycles. 
V0609 Oph displays {\tt RVB} variations, while V0895 Ara, V0967 Cyg, and V0861 Aql are possibly showing a lower degree of this type of variation (i.e., {\tt RVB:}). 
The ASAS-SN classification algorithm performs slightly better for these 55 objects, however, still classifies a majority of the objects as non-RV Tau variables. 
ASAS-SN classified 15 of the 55 objects as RV Tau variable stars. 
Additionally, ASAS-SN classified 19 of the 55 objects as semi-regular and irregular variable stars and 15 objects as Cepheid variables (Type I and Type II, excluding the 15 RV Tau variable stars). 
There are two objects in the saturation regime (V Vul and TW Aql) that ASAS-SN has classified as RV Tau variable stars, however, the alternating minima in the ASAS-SN light curves are faint and not always coherent (see Figure~\ref{fig:rva}).

\subsubsection{Classifications from this work}

Using the classification criteria outlined in \S\ref{classcriteria} and listed in Tables~\ref{tab:rvpgcvs}~and~\ref{tab:rvgcvs}, overall we have identified RV Tau variability in 57 of the 84 objects with irregular RV Tau variability in 8 of these objects. 

The large majority of the known RV Tau in GCVS (i.e., {\tt RV}, {\tt RVA}, {\tt RVB}) exhibit RV Tau variability, as expected. 
For three {\tt RV} or {\tt RVA} stars (BT Lac, V0895 Ara, and V0407 Pav) we identified potential RVB-like variability, possibly for the first time. 
In the ASAS-SN light curves, the {\tt RVB} classification of V1504 Sco is unclear, but its {\tt RVA} classification is clear. 
There are 8 objects whose ASAS-SN light curves do not exhibit clear RV Tau variability, many of these are in the saturation regime and others may reside in crowded fields. 

From the 29 objects with uncertain designations (e.g., {\tt RV:}), 10 show RV Tau variability and we propose a {\tt RVA} classification for these 10 objects. 
Three of these RVA objects have irregularity in the alternating minima (DY Ori, LV Del, and V0576 Aql). 
DZ UMa, which has a {\tt RVB:} designation in GCVS, appears to exhibit RVB-like variability. 
There are 16 objects that are unlikely to be RV Tau variable stars, including V0422 Peg, which is too faint to be considered an RV Tau variable star (see \S\ref{sec:discuss}). 
Three objects remain uncertain or inconclusive (V1071 Cas, V1472 Sgr, and V1377 Sgr). 
Further scrutiny of these uncertain objects is warranted through a separate analysis, and/or with the addition of new data. 

\begin{figure*}
\plotone{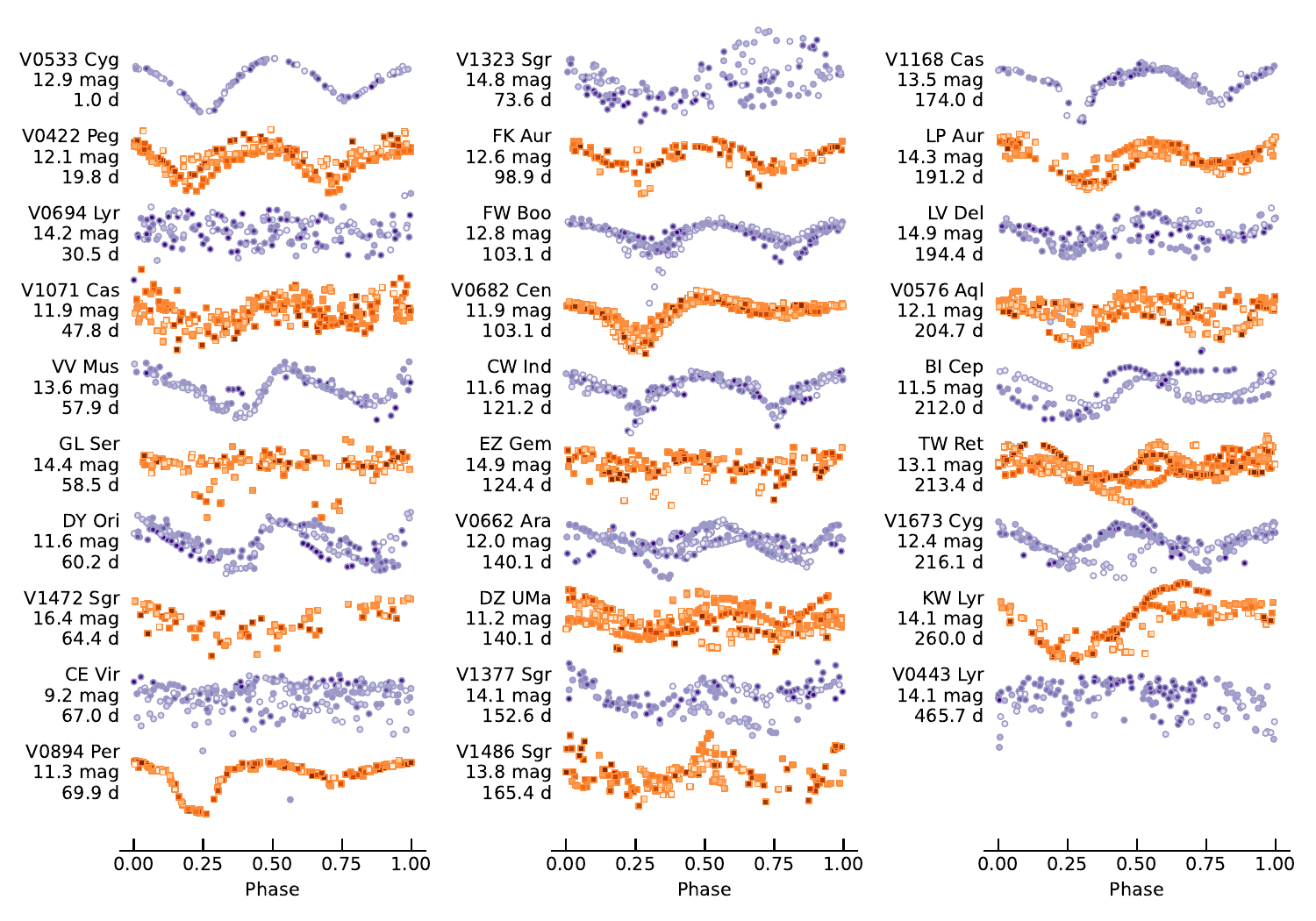}
\caption{Gallery of phase-folded and normalized light curves of stars with the RV:, RVA:, RVB: designation in the GCVS. Curves alternate in color palette and the individual points are color-coded from lighter shades to darker shades by the time of the data point. The star name, median magnitude, and period are indicated to the left of each light curve. The properties of these stars are provided in Table~\ref{tab:rvpgcvs}. 
\label{fig:rvp}}
\end{figure*}

\begin{figure*}
\plotone{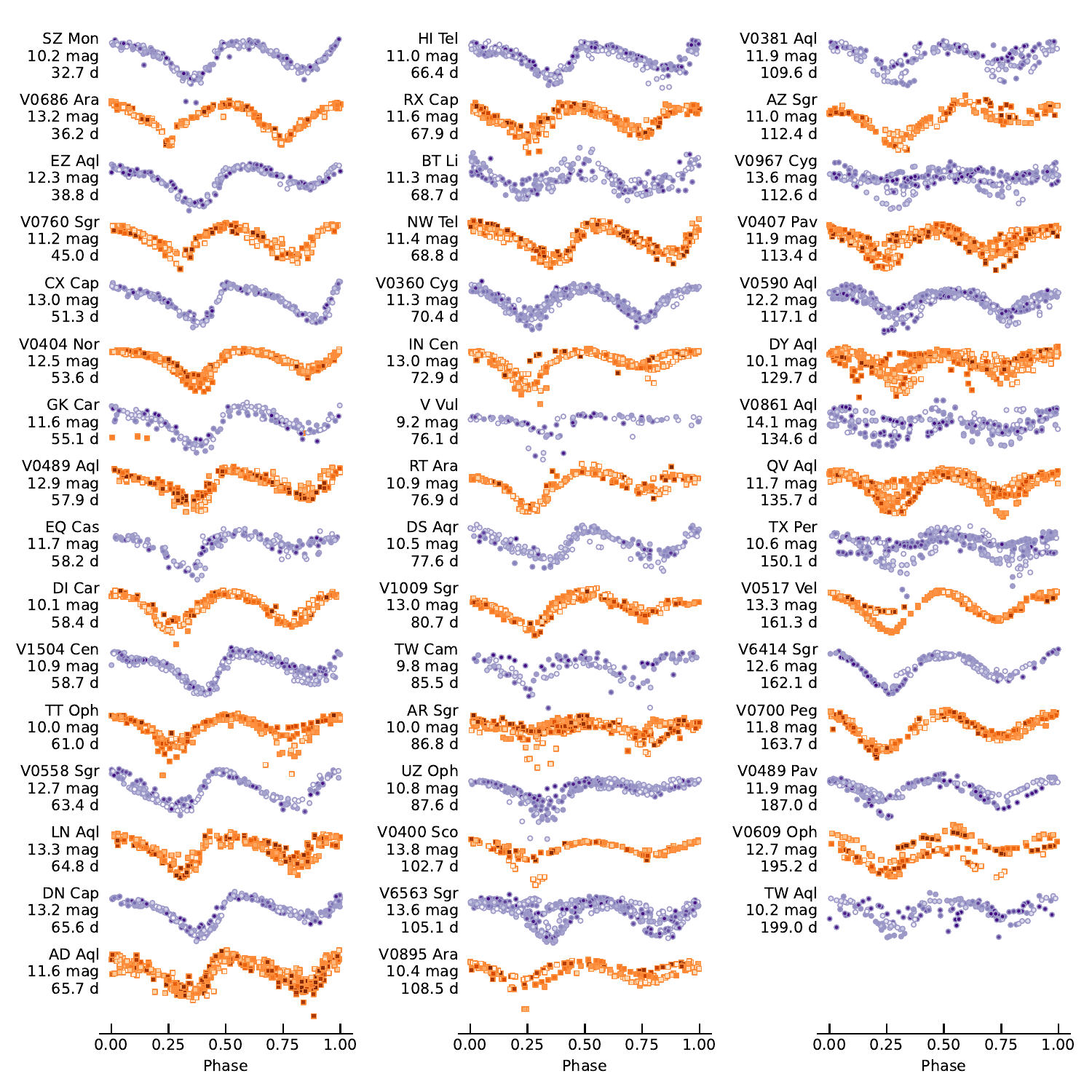}
\caption{Gallery of phase-folded and normalized light curves of stars with the RV and RVA designation in GCVS. The data is further described in the Figure~\ref{fig:rvp} caption.  The properties of these stars are provided in Table~\ref{tab:rvgcvs}. 
\label{fig:rva}}
\end{figure*}

\begin{figure*}
\plotone{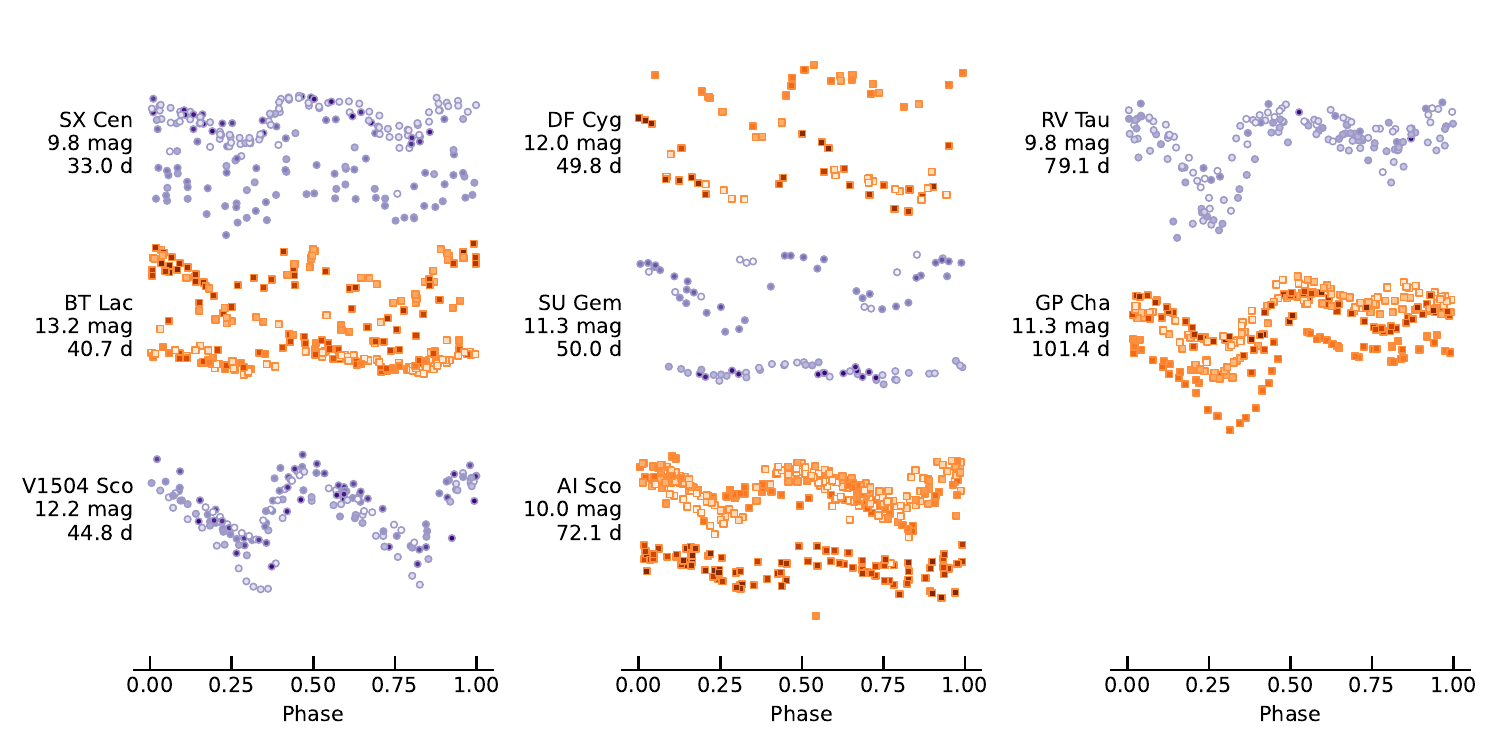}
\caption{Gallery of phase-folded and normalized light curves of stars with the RVB designation in the GCVS. The data is further described in the Figure~\ref{fig:rvp} caption.  The properties of these stars are provided in Table~\ref{tab:rvgcvs}.
\label{fig:rvb}}
\end{figure*}

\section{Discussion} \label{sec:discuss}

Based on our inspection and analysis of ASAS-SN light curves, we have gained insight into the behavior of RV Tau variable stars and challenges for their accurate classification. 
We find that automated classification has difficulty recognizing RV Tau variability due to the alternating minima pattern. 
The periodogram analysis consistently produces the most power in the half-period of the RV Tau formal period. 
In some cases, there is little to no power in the formal period. 
If the peak periodogram period is adopted, then the variable phenomenon is often classified as a shorter period variable, like {\tt DCEP} and {\tt CWA}. 
Visual inspection of phase-folded light curve were necessary to determine if the peak period or its double-period results in the clearest indication of alternating minima. 
This is especially important for those RV Tau variable stars whose minima have similar depths and/or whose deep and shallow minima alternate. 
For periods at or above 150 days, we note that ASAS-SN tends to classify RV Tau variable stars as semi-regular variables, despite the presence of the RV Tau alternating minima pattern.
RV Tau variable stars with long-period changes as the mean magnitude (i.e., {\tt RVB} objects) escape automated classification, however, visual inspection and folding on known periods for the {\tt RVA} variability reveals their RV Tau variability. 
Further analysis of {\tt RVB} objects in the ASAS-SN catalog are suggested. 

These insights of classifying RV Tau variability are complementary to the analysis described in \citet{2019MNRAS.486.1907J}. 
The authors do a similar analysis of doubling the peak period then compare the ratio of the secondary and primary minima at phase $\simeq0.5$ and $\simeq0$ and apply an absolute and ratio threshold to identify RV Tau variables. 
For the long period end, the authors consider infrared colors, amplitude limits, and periods to protect against classifying long period RV Tau variables as semiregular variables. 
Such automated methods are necessary given the volume of sources studied by \citet{2019MNRAS.486.1907J}, but on the scale of our project, visual inspection produces optimal results.
In \citet{2019JAVSO..47..202P} and \citet{2022JAVSO..50...96P}, similar problems with the automated classifications from ASAS-SN were found for semi-regulars, Miras, and other pulsating giants and even for a few RV Tau variables. These authors concluded that ASAS-SN classifications have problems with complex variations from these types of stars.  

Long-term studies like ASAS-SN provide a means to establish the period range of RV Tau variability, which can help test pulsation models \citep{1993A&A...271..501T,1994A&A...292..133F}. 
\citet{1994A&A...292..133F} suggests that non-linear RV Tau pulsation models predict that the shorter period RV Tau variables will be more irregular in comparison to longer periods. 
Long-term photometric studies like that of \citet{1996MNRAS.279..949P}, have considered the regularity of the RV Tau alternating minima variability and suggests that a continuum of behavior ranging from regular variation to irregular may be present amongst the class of RV Tau variable stars.

On the short period end, V0422 Peg exhibits an irregular alternating minima with a period of 19.8 days (Figure~\ref{fig:rvp}), in fact, V0422 Peg is possibly the shortest RV Tau period.
However, calculating the absolute V magnitude using the Gaia parallax suggests $M_{\rm V}\simeq~2~{\rm mag}$, which is too faint for V0422 Peg to be considered an RV Tau variable star. 
A similar conclusion was found for the suspected RV Tau variable star HD~143352 in \citet{2021RNAAS...5...52G} and underscores an important limitation of our study, specifically, we have only considered the light curve variability. 
A subsequent study placing stars into an H-R diagram is necessary and left to future work. 
After V0422 Peg, the next shortest periods in our study belong to SZ Mon, V0686 Ara, and EZ Aql, which all have coherent alternating minima with no evidence of irregularity. 

In our sample, the irregular alternating minima are most abundant in the 100 to 140 day period of the RV Tau deep and shallow minima cycle. 
At the higher period end, several RV Tau variable stars exhibit coherent RV Tau variability above 150 days (e.g., V0517 Vel, V6414 Sgr, and V0700 Peg). 
Meanwhile, many of the suspected RV Tau variable stars with periods above 150 days are more likely to be long period variables (i.e., semi-regular variables). 
The irregularity possibly indicates a relationship with similarly irregular variations seen in SRd variables \citep{2022JAVSO..50...96P}. 
Untangling these interrelationships require long baselines, like those present in the AAVSO archive, such a study would help illuminate any relationship between RV Tau and SRd variables. 
Compared to ASAS-SN baselines, the longer baselines in AAVSO light curves are also important to better determine period variability amongst longer period {\tt RVA} stars and  to identify long-term periods in {\tt RVB} stars.
This underscores how important multi-year observations are in determining the true behavior of the variability.
Among the objects considered in this study, there are some with no clear evidence of alternating minima. 
The lack alternating minima suggests some of these objects are not RV Tau variable stars or their alternating minima have vanished, however, there are saturation and crowded field caveats to consider before making such firm conclusions. 

\section{Conclusions} \label{sec:conclude}

We have identified alternating minima indicative of RV Tau variable stars in 57 of the 84 objects in our study. 
The sample derived from GCVS classifications includes known and suspected RV Tau variable stars. 
For 8 known RV Tau variables, alternating minima are suspect in the ASAS-SN light curves (vanishing or absent). 
Amongst the suspected RV Tau variables, ASAS-SN light curves clearly reveal the presence of alternating minima in 11 stars, thus elevating their RV Tau variable star classification. 
In total we confirmed 47 {\tt RVA} and {\tt RVB} classifications, upgraded 11 suspected variables to {\tt RVA} and {\tt RVB} classifications. 
11 of the stars in our sample have insufficient evidence within ASAS-SN light curves to make conclusive classifications, while 16 are unlikely to be RV Tau variable stars.

By comparing our periods and classifications with those of GCVS and ASAS-SN, we found that the alternating minima phenomenon leads to confusion for ASAS-SN automatic classification. 
Periodograms produce the most power in the half-period, which often does not show alternating mimina when phase-folded. 
As a result, classification algorithms that utilize period ranges may incorrectly classify the objects as shorter period variables. 
Furthermore, this sample of ASAS-SN light curves show the degree of variation that can be present in the alternating minima of RV Tau stars and underscores how important long baseline studies are for identifying and classifying RV Tau stars. 

The contamination of the sample of RV Tau stars has been alluded to in previous studies \citep[e.g.,][]{2019ApJ...872...60B,2021RNAAS...5...52G}. 
We note that stars listed in Simbad with object type {\tt RV*} ({\tt otype=RV*}) are comprised of both known and suspected RV Tau candidates from GCVS ({\tt RV}, {\tt RVA}, and {\tt RVB}, {\tt RV:}, {\tt RVA:}, and {\tt RVB:}). 
As we have shown in our work, those objects with uncertain designations tend to include a larger fraction of stars that do not show clear RV Tau variability. 
Studies that use the Simbad designations are more likely to be contaminated and include objects that may not be RV Tau variables. 
Establishing a robust sample of RV Tau variable stars is an important step towards better understanding the pulsation mechanism and evolutionary path of these variable stars. 
The contribution of surveys like the ASAS-SN and making the data available enables a study like ours and helps bring us closer to a well-vetted large sample of RV Tau variable stars over the largest possible range of periods. 
Combining this data with AAVSO light curves provides a strong tool for characterizing long period and secondary long period phenomenon in RV Tau variable stars. 

\begin{acknowledgments}
This research made use of Astropy, a community-developed core Python package for Astronomy \citep{2018AJ....156..123A, 2013A&A...558A..33A}.

RNN acknowledges support from National Science Foundation under Grant No. 1745460 and the STEM en Route to Change Foundation (SeRCH) Vanguard: Conversations with Women of Color in STEM (\#VanguardSTEM). 

RMJ acknowledges support from NASA contract NAS8-03060.
\end{acknowledgments}

\vspace{5mm}
\facilities{ASAS-SN}

\software{astropy \citep{2013A&A...558A..33A,2018AJ....156..123A}}

\begin{deluxetable*}{lccccccccc}
\tablecaption{RV:, RVA:, and RVB: GCVS Classifications \label{tab:rvpgcvs}}
\tabletypesize{\scriptsize}
\tablehead{
\colhead{Star} & \colhead{VarType} & \colhead{VarType} & \colhead{VarType} & 
\colhead{N} & \colhead{Span} & \colhead{$\overline{V}$} & 
\colhead{$P_{\rm GCVS}$} & \colhead{$P_{\rm ASAS-SN}$} & \colhead{$P_{\rm Adopted}$} \\ 
\colhead{} & \colhead{(GCVS)} & \colhead{(ASAS-SN)} & \colhead{(this work)} & 
\colhead{(points)} & (d) & (mag) &
\colhead{(d)} & \colhead{(d)} & \colhead{(d)} 
}
\startdata
V0533 Cyg & RV: & EB & EB & 91 & 1068 & 12.9 & 63.0 & 1.01 & 1.01 \\
V0422 Peg & RV: & ROT & RVA$^{\dagger}$ & 213 & 1820 & 12.1 & 19.76 & 9.88 & 19.76 \\
V0694 Lyr & RV: & ROT: & NP & 177 & 1363 & 14.2 & 30.52 & \nodata & 30.52 \\
V1071 Cas & RV: & ROT & U & 245 & 1605 & 11.9 & 96.7 & 96.28 & 47.8 \\
VV Mus & RV: & SR & RVA & 163 & 908 & 13.6 & 29.0 & 28.87 & 57.9$\pm$1.6 \\
GL Ser & RV: & SR & NP & 130 & 1325 & 14.4 & \nodata & 29.24 & 58.48 \\
DY Ori & RV: & DCEP & RVA(i) & 184 & 1444 & 11.6 & 60.3 & 30.15 & 60.2$\pm$1.2 \\
V1472 Sgr & RV: & SR & U & 79 & 840 & 16.4 & 69.8 & 43.62 & 64.4 \\
CE Vir & RV: & L & NP$^{\dagger\dagger}$ & 226 & 1725 & 9.2 & 67.0 & \nodata & 67.0 \\
V0894 Per & RV: & CWA & EB & 148 & 1444 & 11.3 & 69.7 & 35.05 & 69.9$\pm$1.4 \\
V1323 Sgr & RV: & SR & SR & 157 & 901 & 14.8 & 61.1 & 86.30 & 73.6 \\
FK Aur & RV: & SR & RVA & 103 & 1440 & 12.6 & 95.9 & 49.20 & 98.9$\pm$3.3 \\
FW Boo & RV: & GCAS & RVA & 228 & 1712 & 12.8 & 102.0 & \nodata & 103.1$\pm$2.4 \\
V0682 Cen & RV: & VAR & RVA & 535 & 1587 & 11.9 & 102.0 & 309.57 & 103.1$\pm$2.4 \\
CW Ind & RV: & GCAS: & RVA & 209 & 1595 & 11.6 & 121.2 & 60.16 & 121.2$\pm$3.1 \\
EZ Gem & RV: & YSO & NP & 154 & 1444 & 14.9 & \nodata & \nodata & 124.4 \\
V0662 Ara & RV: & SR & SR & 318 & 929 & 12.0 & 92.8 & 67.31 & 140.1 \\
DZ UMa & RVB: & SR & RVA/RVB: & 461 & 1854 & 11.2 & \nodata & 70.67 & 140.1$\pm$5.8 \\
V1377 Sgr & RV: & SR & U & 157 & 901 & 14.1 & 60.0 & 78.10 & 152.6$\pm$13.2 \\
V1486 Sgr & RV: & SR & SR & 157 & 901 & 13.8 & 75.0 & 80.65 & 165.4$\pm$13.4 \\
V1168 Cas & RV: & SR & EB & 130 & 1412 & 13.5 & 172.0 & 174.00 & 174.0$\pm$8.4 \\
LP Aur & RV: & SR & RVA & 217 & 1775 & 14.3 & \nodata & 97.75 & 191.2$\pm$8.1 \\
LV Del & RV: & SR & RVA(i) & 169 & 1689 & 14.9 & \nodata & 97.22 & 194.44 \\
V0576 Aql & RVA: & L & RVA(i) & 248 & 1694 & 12.1 & 201.2 & \nodata & 204.7 \\
BI Cep & RVA: & SR & SR: & 153 & 1103 & 11.5 & 212.0 & 107.09 & 212.0 \\
TW Ret & RV: & SR & SR: & 382 & 1553 & 13.1 & \nodata & 213.37 & 213.37 \\
V1673 Cyg & RV: & SR & SR: & 269 & 1715 & 12.4 & \nodata & 108.04 & 216.08 \\
KW Lyr & RV: & SR & SR: & 174 & 1588 & 14.1 & 260.0 & 273.86 & 260.0 \\
V0443 Lyr & RV: & YSO & NP & 174 & 1588 & 14.1 & \nodata & \nodata & 465.7 \\
\enddata
\tablenotetext{\dagger}{ As discussed in the text, V0422 Peg is too faint to be considered an RV Tau variable star.}
\tablenotetext{\dagger\dagger}{ Although no periodicity is found for CE Vir, we note this star is in the saturation regime.}
\tablecomments{VarType is the variable type identified in the GCVS and ASAS-SN catalogs and this work. N, Span, and $\overline{V}$ are the number of data points, span of the observations, and mean magnitude based on the ASAS-SN light curves. For the VarType from this work, we use ``(i)'' to indicate when alternating minima are irregular. }
\end{deluxetable*}

\begin{deluxetable*}{lccccccccc}
\tablecaption{RV, RVA, and RVB GCVS Classifications \label{tab:rvgcvs}}
\tablewidth{700pt}
\tablehead{
\colhead{Star} & \colhead{VarType} & \colhead{VarType} & \colhead{VarType} & 
\colhead{N} & \colhead{Span} &  \colhead{$\overline{V}$} & 
\colhead{$P_{\rm GCVS}$} & \colhead{$P_{\rm ASAS-SN}$} & \colhead{$P_{\rm Adopted}$} \\ 
\colhead{} & \colhead{(GCVS)} & \colhead{(ASAS-SN)} & \colhead{(this work)} & 
\colhead{(points)} & (d) & (mag) & 
\colhead{(d)} & \colhead{(d)} & \colhead{(d)} 
}
\startdata
SZ Mon & RVA & DCEP & RVA$^{\dagger}$ & 265 & 1367 & 10.2 & 32.7 & 16.34 & 32.7$\pm$0.3 \\
SX Cen & RVB & L & RVB$^{\dagger\dagger}$ & 204 & 916 & 9.8 & 33.0 & \nodata & 33.0 \\
V0686 Ara & RV & VAR & RVA & 263 & 928 & 13.2 & 36.3 & \nodata & 36.2$\pm$0.5 \\
EZ Aql & RVA & RVA & RVA & 205 & 1299 & 12.3 & 38.7 & 38.65 & 38.8$\pm$0.9 \\
BT Lac & RVB & YSO & RVB & 228 & 1094 & 13.2 & 40.5 & \nodata & 40.7$\pm$0.7 \\
V1504 Sco & RVB & DCEP & RVA/RVB: & 163 & 929 & 12.2 & 44.0 & 22.39 & 44.8$\pm$0.8 \\
V0760 Sgr & RVA & DCEP & RVA & 164 & 1312 & 11.2 & 45.3 & 22.51 & 45.0$\pm$0.5 \\
DF Cyg & RVB & M: & RVB & 69 & 990 & 12.0 & 49.8 & \nodata & 49.8 \\
SU Gem & RVB & M: & RVB & 94 & 1040 & 11.3 & 50.0 & \nodata & 50.0 \\
CX Cap & RVA & CWA & RVA & 204 & 1584 & 13.0 & 51.4 & 25.64 & 51.3$\pm$0.6 \\
V0404 Nor & RVA & RVA & RVA & 364 & 930 & 12.5 & 53.6 & 53.91 & 53.6$\pm$1.3 \\
GK Car & RVA & RVA & RVA & 172 & 913 & 11.6 & 55.3 & 55.15 & 55.1$\pm$1.4 \\
V0489 Aql & RVA & CWA & RVA & 307 & 1328 & 12.9 & 57.9 & 28.88 & 57.9$\pm$1.0 \\
EQ Cas & RVA & CWA & RVA & 160 & 1441 & 11.7 & 58.3 & 58.52 & 58.2$\pm$0.9 \\
DI Car & RV & L & RVA$^{\dagger}$ & 189 & 895 & 10.1 & 58.3 & \nodata & 58.4$\pm$1.5 \\
V1504 Cen & RVA & RVA & RVA$^{\dagger}$ & 285 & 1571 & 10.9 & 58.7 & 58.67 & 58.7$\pm$0.8 \\
TT Oph & RVA & VAR & RVA$^{\dagger}$ & 232 & 1953 & 10.0 & 61.1 & \nodata & 61.0$\pm$0.9 \\
V0558 Sgr & RV & CWA & RVA & 183 & 921 & 12.7 & 63.8 & 31.72 & 63.4$\pm$1.5 \\
LN Aql & RVA & CWA & RVA & 315 & 1362 & 13.3 & 64.7 & 48.66 & 64.8$\pm$1.2 \\
DN Cap & RVA & RVA & RVA & 260 & 1635 & 13.2 & 65.5 & 131.25 & 65.6$\pm$2.1 \\
AD Aql & RVA & DCEP & RVA & 424 & 1168 & 11.6 & 66.0 & 32.88 & 65.7$\pm$1.3 \\
HI Tel & RVA & RVA & RVA$^{\dagger}$ & 277 & 1591 & 11.0 & 66.5 & 66.35 & 66.4$\pm$0.9 \\
RX Cap & RVA & CWA & RVA & 237 & 1976 & 11.6 & 67.9 & 33.97 & 67.9$\pm$0.9 \\
BT Lib & RV & CWA & RVA/RVB: & 230 & 1697 & 11.3 & 75.3 & 34.34 & 68.7$\pm$1.2 \\
NW Tel & RVA & RVA & RVA & 261 & 1590 & 11.4 & 68.9 & 68.70 & 68.8$\pm$1.0 \\
V0360 Cyg & RVA & CWA & RVA & 378 & 1854 & 11.3 & 70.4 & 35.22 & 70.4$\pm$1.3 \\
AI Sco & RVB & M: & RVB$^{\dagger}$ & 344 & 928 & 10.0 & 71.0 & \nodata & 72.1$\pm$1.9 \\
IN Cen & RVA & SR & RVA & 206 & 937 & 13.0 & 75.8 & 36.84 & 72.9$\pm$6.6 \\
V Vul & RVA & RVA & U$^{\dagger\dagger}$ & 86 & 936 & 9.2 & 75.7 & 75.78 & 76.1$\pm$4.8 \\
RT Ara & RVA & RVA & RVA$^{\dagger}$ & 164 & 927 & 10.9 & 76.7 & 77.10 & 76.9$\pm$5.8 \\
DS Aqr & RVA & RVA & RVA$^{\dagger}$ & 189 & 1751 & 10.5 & 77.3 & 77.49 & 77.6$\pm$1.9 \\
RV Tau & RVB & L & RVB$^{\dagger\dagger}$ & 157 & 2500 & 9.8 & 78.7 & \nodata & 79.1$\pm$1.5 \\
V1009 Sgr & RVA & RVA & RVA & 183 & 921 & 13.0 & 80.0 & 81.08 & 80.7$\pm$2.3 \\
TW Cam & RV & CWA & U$^{\dagger\dagger}$ & 128 & 1443 & 9.8 & 86.0 & 43.14 & 85.5$\pm$2.3 \\
AR Sgr & RVA & CWA & RVA$^{\dagger}$ & 319 & 1641 & 10.0 & 87.9 & 43.40 & 86.8$\pm$1.5 \\
UZ Oph & RVA & RVA & RVA$^{\dagger}$ & 281 & 1706 & 10.8 & 87.4 & 87.50 & 87.6$\pm$4.1 \\
GP Cha & RVB & L & RVB & 246 & 1530 & 11.3 & 100.6 & \nodata & 101.4$\pm$6.1 \\
V0400 Sco & RV & SR & U(i) & 115 & 740 & 13.8 & 102.8 & 51.39 & 102.7$\pm$4.8 \\
V6563 Sgr & RVA & VAR & RVA(i) & 495 & 1596 & 13.6 & 105.1 & 158.12 & 105.1$\pm$2.8 \\
V0895 Ara & RVA & L & RVA(i)/RVB:$^{\dagger}$ & 143 & 925 & 10.4 & 108.7 & \nodata & 108.5$\pm$5.4 \\
V0381 Aql & RVA & SR & RVA(i) & 194 & 1321 & 11.9 & 109.6 & 55.15 & 109.6$\pm$3.6 \\
AZ Sgr & RVA & RVA & RVA(i)$^{\dagger}$ & 143 & 923 & 11.0 & 113.6 & 111.82 & 112.4$\pm$10.0 \\
V0967 Cyg & RV & SR & U & 332 & 1312 & 13.6 & 50.0 & 56.29 & 112.59 \\
V0407 Pav & RV & RVA & RVA/RVB: & 371 & 1406 & 11.9 & 112.8 & 113.37 & 113.4$\pm$3.31 \\
V0590 Aql & RVA & CWA & RVA & 579 & 1694 & 12.2 & 117.4 & 58.47 & 117.1$\pm$3.1 \\
DY Aql & RVA & L & U(i)$^{\dagger}$ & 439 & 1686 & 10.1 & 131.9 & \nodata & 129.7$\pm$9.3 \\
V0861 Aql & RV & SR & U & 339 & 1353 & 14.1 & 110.6 & 67.29 & 134.57 \\
QV Aql & RVA & SR & RVA(i) & 426 & 1599 & 11.7 & 135.0 & 67.77 & 135.7$\pm$5.3 \\
TX Per & RVA & L & U$^{\dagger}$ & 413 & 1833 & 10.6 & 78.0 & \nodata & 150.1$\pm$8.8 \\
V0517 Vel & RVA & SR & RVA & 304 & 774 & 13.3 & 160.8 & 80.62 & 161.3$\pm$15.7 \\
V6414 Sgr & RVA & SR & RVA & 193 & 927 & 12.6 & 161.3 & 80.87 & 162.1$\pm$10.1 \\
V0700 Peg & RVA & SR & RVA & 217 & 1853 & 11.8 & 163.7 & 81.99 & 163.7$\pm$5.6 \\
V0489 Pav & RVA & SR & RVA & 201 & 1600 & 11.9 & 188.0 & 94.07 & 187.0$\pm$8.2 \\
V0609 Oph & RV & SR & RVB & 153 & 1338 & 12.7 & 195.0 & 99.25 & 195.2$\pm$11.2 \\
TW Aql & RVA & RVA & U$^{\dagger}$ & 164 & 1117 & 10.2 & 179.0 & 100.47 & 199.0$\pm$26.8 \\
\enddata
\tablenotetext{\dagger}{ Indicates a star in the 10-11 mag saturation regime.}
\tablenotetext{\dagger\dagger}{ Indicates a star in the $<$10 mag saturation regime.}
\tablecomments{VarType is the variable type identified in the GCVS and ASAS-SN catalogs and this work. N, Span, and $\overline{V}$ are the number of data points, span of the observations, and mean magnitude based on the ASAS-SN light curves. For the VarType from this work, we use ``(i)'' to indicate when alternating minima are irregular. }
\end{deluxetable*}


\bibliography{nerervtau}{}

\begin{thebibliography}{}
\expandafter\ifx\csname natexlab\endcsname\relax\def\natexlab#1{#1}\fi
\providecommand{\url}[1]{\href{#1}{#1}}
\providecommand{\dodoi}[1]{doi:~\href{http://doi.org/#1}{\nolinkurl{#1}}}
\providecommand{\doeprint}[1]{\href{http://ascl.net/#1}{\nolinkurl{http://ascl.net/#1}}}
\providecommand{\doarXiv}[1]{\href{https://arxiv.org/abs/#1}{\nolinkurl{https://arxiv.org/abs/#1}}}

\bibitem[{{Astropy Collaboration} {et~al.}(2013){Astropy Collaboration},
  {Robitaille}, {Tollerud}, {Greenfield}, {Droettboom}, {Bray}, {Aldcroft},
  {Davis}, {Ginsburg}, {Price-Whelan}, {Kerzendorf}, {Conley}, {Crighton},
  {Barbary}, {Muna}, {Ferguson}, {Grollier}, {Parikh}, {Nair}, {Unther},
  {Deil}, {Woillez}, {Conseil}, {Kramer}, {Turner}, {Singer}, {Fox}, {Weaver},
  {Zabalza}, {Edwards}, {Azalee Bostroem}, {Burke}, {Casey}, {Crawford},
  {Dencheva}, {Ely}, {Jenness}, {Labrie}, {Lim}, {Pierfederici}, {Pontzen},
  {Ptak}, {Refsdal}, {Servillat}, \& {Streicher}}]{2013A&A...558A..33A}
{Astropy Collaboration}, {Robitaille}, T.~P., {Tollerud}, E.~J., {et~al.} 2013,
  \aap, 558, A33, \dodoi{10.1051/0004-6361/201322068}

\bibitem[{{Astropy Collaboration} {et~al.}(2018){Astropy Collaboration},
  {Price-Whelan}, {Sip{\H{o}}cz}, {G{\"u}nther}, {Lim}, {Crawford}, {Conseil},
  {Shupe}, {Craig}, {Dencheva}, {Ginsburg}, {VanderPlas}, {Bradley},
  {P{\'e}rez-Su{\'a}rez}, {de Val-Borro}, {Aldcroft}, {Cruz}, {Robitaille},
  {Tollerud}, {Ardelean}, {Babej}, {Bach}, {Bachetti}, {Bakanov}, {Bamford},
  {Barentsen}, {Barmby}, {Baumbach}, {Berry}, {Biscani}, {Boquien}, {Bostroem},
  {Bouma}, {Brammer}, {Bray}, {Breytenbach}, {Buddelmeijer}, {Burke},
  {Calderone}, {Cano Rodr{\'\i}guez}, {Cara}, {Cardoso}, {Cheedella}, {Copin},
  {Corrales}, {Crichton}, {D'Avella}, {Deil}, {Depagne}, {Dietrich}, {Donath},
  {Droettboom}, {Earl}, {Erben}, {Fabbro}, {Ferreira}, {Finethy}, {Fox},
  {Garrison}, {Gibbons}, {Goldstein}, {Gommers}, {Greco}, {Greenfield},
  {Groener}, {Grollier}, {Hagen}, {Hirst}, {Homeier}, {Horton}, {Hosseinzadeh},
  {Hu}, {Hunkeler}, {Ivezi{\'c}}, {Jain}, {Jenness}, {Kanarek}, {Kendrew},
  {Kern}, {Kerzendorf}, {Khvalko}, {King}, {Kirkby}, {Kulkarni}, {Kumar},
  {Lee}, {Lenz}, {Littlefair}, {Ma}, {Macleod}, {Mastropietro}, {McCully},
  {Montagnac}, {Morris}, {Mueller}, {Mumford}, {Muna}, {Murphy}, {Nelson},
  {Nguyen}, {Ninan}, {N{\"o}the}, {Ogaz}, {Oh}, {Parejko}, {Parley}, {Pascual},
  {Patil}, {Patil}, {Plunkett}, {Prochaska}, {Rastogi}, {Reddy Janga},
  {Sabater}, {Sakurikar}, {Seifert}, {Sherbert}, {Sherwood-Taylor}, {Shih},
  {Sick}, {Silbiger}, {Singanamalla}, {Singer}, {Sladen}, {Sooley},
  {Sornarajah}, {Streicher}, {Teuben}, {Thomas}, {Tremblay}, {Turner},
  {Terr{\'o}n}, {van Kerkwijk}, {de la Vega}, {Watkins}, {Weaver}, {Whitmore},
  {Woillez}, {Zabalza}, \& {Astropy Contributors}}]{2018AJ....156..123A}
{Astropy Collaboration}, {Price-Whelan}, A.~M., {Sip{\H{o}}cz}, B.~M., {et~al.}
  2018, \aj, 156, 123, \dodoi{10.3847/1538-3881/aabc4f}

\bibitem[{{B{\'o}di} \& {Kiss}(2019)}]{2019ApJ...872...60B}
{B{\'o}di}, A., \& {Kiss}, L.~L. 2019, \apj, 872, 60,
  \dodoi{10.3847/1538-4357/aafc24}

\bibitem[{{Fokin}(1994)}]{1994A&A...292..133F}
{Fokin}, A.~B. 1994, \aap, 292, 133

\bibitem[{{Graber} \& {Montez}(2021)}]{2021RNAAS...5...52G}
{Graber}, S., \& {Montez}, Rodolfo, J. 2021, Research Notes of the American
  Astronomical Society, 5, 52, \dodoi{10.3847/2515-5172/abf046}

\bibitem[{{Jayasinghe} {et~al.}(2019){Jayasinghe}, {Stanek}, {Kochanek},
  {Shappee}, {Holoien}, {Thompson}, {Prieto}, {Dong}, {Pawlak}, {Pejcha},
  {Shields}, {Pojmanski}, {Otero}, {Britt}, \& {Will}}]{2019MNRAS.486.1907J}
{Jayasinghe}, T., {Stanek}, K.~Z., {Kochanek}, C.~S., {et~al.} 2019, \mnras,
  486, 1907, \dodoi{10.1093/mnras/stz844}

\bibitem[{{Jayasinghe} {et~al.}(2020){Jayasinghe}, {Stanek}, {Kochanek},
  {Shappee}, {Holoien}, {Thompson}, {Prieto}, {Dong}, {Pawlak}, {Pejcha},
  {Shields}, {Pojmanski}, {Otero}, {Hurst}, {Britt}, \&
  {Will}}]{2020MNRAS.491...13J}
---. 2020, \mnras, 491, 13, \dodoi{10.1093/mnras/stz2711}

\bibitem[{{Kochanek} {et~al.}(2017){Kochanek}, {Shappee}, {Stanek}, {Holoien},
  {Thompson}, {Prieto}, {Dong}, {Shields}, {Will}, {Britt}, {Perzanowski}, \&
  {Pojma{\'n}ski}}]{2017PASP..129j4502K}
{Kochanek}, C.~S., {Shappee}, B.~J., {Stanek}, K.~Z., {et~al.} 2017, \pasp,
  129, 104502, \dodoi{10.1088/1538-3873/aa80d9}

\bibitem[{{Lomb}(1976)}]{1976Ap&SS..39..447L}
{Lomb}, N.~R. 1976, \apss, 39, 447, \dodoi{10.1007/BF00648343}

\bibitem[{{Payne-Gaposchkin} {et~al.}(1943){Payne-Gaposchkin}, {Brenton}, \&
  {Gaposchkin}}]{1943AnHar.113....1P}
{Payne-Gaposchkin}, C., {Brenton}, V.~K., \& {Gaposchkin}, S. 1943, Annals of
  Harvard College Observatory, 113, 1

\bibitem[{{Percy}(2022)}]{2022JAVSO..50...96P}
{Percy}, J.~R. 2022, \jaavso, 50, 96

\bibitem[{{Percy} \& {Fenaux}(2019)}]{2019JAVSO..47..202P}
{Percy}, J.~R., \& {Fenaux}, L. 2019, \jaavso, 47, 202,
  \dodoi{10.48550/arXiv.1905.03279}

\bibitem[{{Percy} {et~al.}(1991){Percy}, {Sasselov}, {Alfred}, \&
  {Scott}}]{1991ApJ...375..691P}
{Percy}, J.~R., {Sasselov}, D.~D., {Alfred}, A., \& {Scott}, G. 1991, \apj,
  375, 691, \dodoi{10.1086/170233}

\bibitem[{{Pollard} {et~al.}(1996){Pollard}, {Cottrell}, {Kilmartin}, \&
  {Gilmore}}]{1996MNRAS.279..949P}
{Pollard}, K.~R., {Cottrell}, P.~L., {Kilmartin}, P.~M., \& {Gilmore}, A.~C.
  1996, \mnras, 279, 949, \dodoi{10.1093/mnras/279.3.949}

\bibitem[{{Samus'} {et~al.}(2017){Samus'}, {Kazarovets}, {Durlevich},
  {Kireeva}, \& {Pastukhova}}]{2017ARep...61...80S}
{Samus'}, N.~N., {Kazarovets}, E.~V., {Durlevich}, O.~V., {Kireeva}, N.~N., \&
  {Pastukhova}, E.~N. 2017, Astronomy Reports, 61, 80,
  \dodoi{10.1134/S1063772917010085}

\bibitem[{{Scargle}(1982)}]{1982ApJ...263..835S}
{Scargle}, J.~D. 1982, \apj, 263, 835, \dodoi{10.1086/160554}

\bibitem[{{Shappee} {et~al.}(2014){Shappee}, {Prieto}, {Grupe}, {Kochanek},
  {Stanek}, {De Rosa}, {Mathur}, {Zu}, {Peterson}, {Pogge}, {Komossa}, {Im},
  {Jencson}, {Holoien}, {Basu}, {Beacom}, {Szczygie{\l}}, {Brimacombe},
  {Adams}, {Campillay}, {Choi}, {Contreras}, {Dietrich}, {Dubberley},
  {Elphick}, {Foale}, {Giustini}, {Gonzalez}, {Hawkins}, {Howell}, {Hsiao},
  {Koss}, {Leighly}, {Morrell}, {Mudd}, {Mullins}, {Nugent}, {Parrent},
  {Phillips}, {Pojmanski}, {Rosing}, {Ross}, {Sand}, {Terndrup}, {Valenti},
  {Walker}, \& {Yoon}}]{2014ApJ...788...48S}
{Shappee}, B.~J., {Prieto}, J.~L., {Grupe}, D., {et~al.} 2014, \apj, 788, 48,
  \dodoi{10.1088/0004-637X/788/1/48}

\bibitem[{{Tuchman} {et~al.}(1993){Tuchman}, {Lebre}, {Mennessier}, \&
  {Yarri}}]{1993A&A...271..501T}
{Tuchman}, Y., {Lebre}, A., {Mennessier}, M.~O., \& {Yarri}, A. 1993, \aap,
  271, 501

\bibitem[{{Wallerstein}(2002)}]{2002PASP..114..689W}
{Wallerstein}, G. 2002, \pasp, 114, 689, \dodoi{10.1086/341698}

\bibitem[{{Zsoldos}(1991)}]{1991IBVS.3557....1Z}
{Zsoldos}, E. 1991, Information Bulletin on Variable Stars, 3557, 1

\end{thebibliography}
\bibliographystyle{aasjournal}

\end{document}